# Cortical representations of Auditory Perception using Graph Independent Component Analysis on EEG

Pranav Sankhe[1†]
Ritik Madan[1]
[1] Indian Institute of Technology Bombay, Mumbai India
† Corresponding author: pranavsankhe40@gmail.com

## Abstract

Recent studies indicate that the neurons involved in a cognitive task aren't locally limited but span out to multiple human brain regions. We obtain network components and their locations for the task of listening to music. The recorded EEG data is modeled as a graph, and it is assumed that the overall activity is a contribution of several independent subnetworks. To identify these intrinsic cognitive subnetworks corresponding to music perception, we propose to decompose the whole brain graph-network into multiple subnetworks. We perform this decomposition to a group of brain networks by performing Graph-Independent Component Analysis. Graph-ICA is a variant of ICA that decomposes the measured graph into independent source graphs. Having obtained independent subnetworks, we calculate the electrode positions by computing the local maxima of these subnetwork matrices. We observe that the computed electrodes' location corresponds to the temporal lobes and the Broca's area, which are indeed involved in the task of auditory processing and perception. The computed electrodes also span the brain's frontal lobe, which is involved in attention and generating a stimulus-evoked response. The weight of the subnetwork that corresponds to the aforementioned brain regions increases with the increase in the music recording's tempo. The results suggest that whole-brain networks can be decomposed into independent subnetworks and analyze cognitive responses to music stimulus.

## Introduction

One of the most remarkable aspects of the human brain is its ability to recognize patterns and describe them. Among the amazing patterns we've tried to understand, music remains to be the most intensely thought and studied about. The process of studying ubiquitous properties of musical patterns is challenging because the interpretation of music has been inherently subjective through human history.

Our body carries out several varieties of tasks and almost all of those are carried directly or indirectly by the brain. An important question to ask is how neurons are allotted to a task and what is the relationship between neurons allotted to a particular task. This relationship can be modeled as a graph. From the neuroscientific studies of Bullmore (2009) and Bressler et al. (2010), we know that there exists a structure to function relationship for every cognition task. We consider this structural relationship from a network perspective. Studies show that there are a finite set of sub-networks which are dynamically allotted to cognitive tasks like listening to music.

There are many ways to study music perception. In our approach, we model brain activity as a graph. The hypothesis is that there are few underlying subnetworks that contribute to the overall functioning of the brain. To find out these underlying subnetworks, we perform Independent Component Analysis ICA on the EEG data recorded when the subjects were listening to music. Having obtained the subnetworks and thus the position of the electrodes, we intend to correspond them to the areas which according to literature, are indeed the regions who do the auditory processing.

## Method

*Dataset*
We use Stanford's NMED-T dataset containing cortical (EEG) and behavioral data collected during natural music listening by Kaneshiro et al. (2017). EEG signals were recorded from 20 adult participants while listening to a set of 10 full-length songs with electronically produced beats at various tempos. The sampling rate of the EEG recording is 125Hz. The EEG data were pre-processed to remove artifacts.

*Approach: Graph ICA (G-ICA)*
To understand ICA, consider the cocktail party problem. Here, n speakers are speaking simultaneously at a party, and any microphone placed in the room records only an overlapping combination of the n speakers' voices. Let us say we have n different microphones placed in the room, and because each microphone is a different distance from each of the speakers, it records a different combination of the speakers' voices. We ask whether using the recordings from microphones, can we separate out the original n speakers' speech signals? Independent Component Analysis essentially answers this question with a resounding yes.

Our approach is inspired by Park et al. (2014)'s approach to study networks using fMRI data for particular cognitive tasks. Graph-ICA is a variant of ICA that decomposes measured graphs into independent source graphs. We represent the graph by its adjacency matrix. We denote a graph (i.e. an adjacency matrix) with L nodes of the i-th brain, a vector, $g_i$, with K = L(L−1)/2 edges for elements. Assume N independent network components $s_j$, j = 1, .., N exist in the brain. M graphs from M brains, i.e., $g_i$, i = 1, ..., M were concatenated to a matrix **g**.

$$\mathbf{g_i} = \sum_{j=1}^{N} \mathbf{a_{ij} s_i}$$

where, $\mathbf{a_{ij}}$ is the element of mixing matrix **A** indicating the contribution of (graph) source $\mathbf{s_j}$.

For this study, we assumed the number of ICs (N) equaled the number of graphs (M), i.e., N = M = 20, since we do not have a clear a priori knowledge of the number of ICs. The mixing matrix A can be estimated by an ICA algorithm, which maximizes mutual independence between estimated functional components.

For each participant, only upper diagonal elements of the adjacency matrix were used for graph-ICA since the adjacency matrix is symmetric in this study. The upper diag- onal elements were vectorised. The subnetwork matrices obtained from ICA are also vectorised. To convert the vectorised subnetworks into matrix form, the vector is considered as a representative of the upper diagonal matrix. We employ ICA over these vectors. Having obtained the subnetworks, we identify the active EEG electrodes corresponding to the individual subnetworks and plot them on the surface of an inflated brain in order to study the cortical areas they correspond to.

## Results

Once we get all the 20 subnetworks, we rank them based on their weights in the mixture. We observe that the subnetwork with the second-highest weight corresponds to the cortical regions, which are indeed responsible for audio processing in our brain. This observation does not change when we change the music stimuli (thus changing the tempo), which indicates the validity of the results.

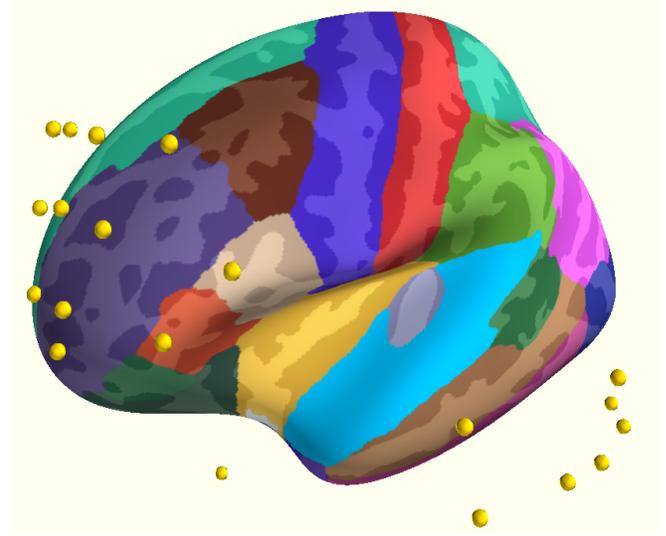

*Figure 1: Subnetwork electrodes corresponding to auditory perception cortical areas*

We observed that the subnetwork corresponding to auditory cortical areas covered in particular the Rostral Middle Frontal, Superior Frontal, Pars Opercularis, Pars Triangularis, Superior Temporal, and Transverse Temporal region of the brain. The Pars Opercularis, Pars Triangulairis regions are involved in semantic understanding of audio stimuli according to the study by Maess et al. (2001). The Superior Temporal and Transverse Temporal cortical areas belong to the auditory cortex, which is involved in the cognitive processing of incoming audio stimuli as found by Pickles at al (2012). The frontal cortical area carries out functions like coordinating sensory systems, emotion, and carrying out tasks involving past memory as reported by Goldberg et al. (2006).

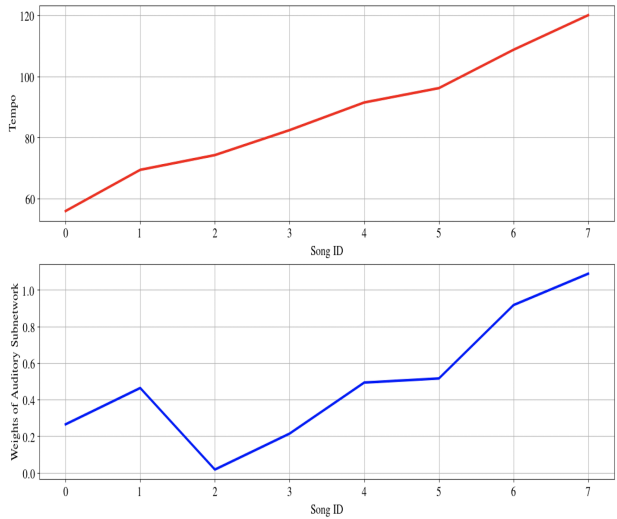

*Figure 2: Subplot 1 shows how tempo varies with the audio stimuli. Subplot 2 depicts the variation of weight of auditory subnetwork with the audio stimuli*

After identifying the auditory subnetwork, we studied how its weight (obtained by G-ICA) varies with the tempo of the music stimuli. We observed that in almost all the cases, the weight of the auditory subnetwork increases proportionately with the tempo of the stimuli.

## Conclusion

We applied Graph Independent Component Analysis on EEG data to study brain activity networks for music perception. We identified the auditory subnetwork and studied its variation with the tempo of the music stimuli. The proportional relationship between the tempo and the weight of the subnetwork is encouraging for further investigation of different properties of music stimuli and understanding how they correspond with the subnetwork.